# Quantifying socioeconomic activities and weather effects on the global spread of COVID-19 epidemic

Xinyi Shen[1*], Chenkai Cai[1,2], and Hui Li[3]

[1] Department of Civil and Environmental Engineering, University of Connecticut, Storrs, CT, 06269

[2] College of Hydrology and Water Resources, Hohai University, Nanjing, China, 210098

[3] Department of Finance, University of Connecticut, Storrs, CT, 06269

## Summary

The COVID-19 has caused more than three million infections and over two hundred thousand deaths by April 2020[1]. Limiting socioeconomic activities (SA) is among the most adopted governmental mitigating efforts to combat the transmission of the virus, though the degree varies dramatically among different regimes[2]. This study aims to quantify the contribution from the SA and weather conditions to the transmission of COVID-19 at global scale. Ruling out the unobservable factors including medical facilities and other control policies (MOC) through region-by-time fixed effects[3,4], we show that the limiting SA has a leading contribution to lower the reproductive number by 18.3%, while weather conditions, including ultraviolet, relative humidity, and wind explain a smaller amount of variation. Temperature might have a non-monotonic impact on the transmission. We further show that in developed countries[5] and China, the SA effect is more pronounced whereas the weather effect is significantly downplayed possibly because people tend to stay indoors most of the time with a controlled climate. We finally estimate the reduced reproductive number and the population spared from infections due to restricting SA at 40,964, 180,336, 174,494, in China, United States, and Europe respectively. From late January to mid-April, all regions, except for China, Australia, and south Korea show a steep upward trend of spared infections due to restricting SA. US and Europe, in particular, show far steeper upward trends of spared infections in the analyzed timeframe, signaling a greater risk of reopening the economy too soon.

## Main Text

The COVID-19, declared global pandemic by the World Health Organization (WHO), is reporeted to have infected more than three million population in 213 countries and territories by April 2020[1]. Up-to-date, it has a 30-40 times death rate of seasonal influenza (3-4% vs. 0.1%) and infects all ages[2]. Managing a controlled transition from community transmission to a low-level or no transmission is, at present, the best-case outcome in the short and medium term in the absence of an effective treatment and vaccine[6]. The restriction of socioeconomic activity (SA) is a widely adopted mitigation management by national and subnational authorities for combating the virus. Strong form of restrictions involves city lock down, mandatory quarantine, and large-scale demobilization while weak form includes shelter-in-place order allowing for the openness of essential businesses and domestic travel. The degree to which SA is restricted varies dramatically with location, time, and climate regions, and is difficult to measure directly. Additional difficulty lies with civilians' compliance with governmental orders. Consequently, the impact on suppressing the COVID-19 transmissibility of limiting SA in various weather conditions remains unknown[7].

Previous studies on restriction effects rely on news of governmental announcements controls[8-11] or human travel records[12] to predict the number of infectees in a single city, country, or regime[12-15] based on the stochastic susceptible-exposed-infectious removed (SEIR) modeling framework or its variance.

Hypothetical scenarios of lifting the restrictions can therefore be estimated after SEIR is parameterized. These methods are, however, difficulty to applied globally because: 1) human travel records obtained from a personal tracking system do not exist in most countries, 2) the announcement of control policies and the practical implementation and public compliance can be inconsistent, and 3) a process model like SEIR requires many regionally heterogeneous epidemic parameters[11,12] which are unavailable globally. Furthermore, validating the model only by the records from a strictly controlled region or country (e.g. Wuhan or China) might bias the estimation of an extreme scenario with no controls. Without using the global data, it is also difficult to assess the seasonal effects.

We build a panel regression model (Methods) to quantify contribution of SA, climate, and MOC to the transmissibility of COVID-19 over the globe during the community spread period. We estimate the daily reproductive number, $R_0$, of COVID-19, following a widely adopted statistical approach[8,16] which only requires the daily confirmed number of infectees[1]. $R_0$ is an indicator of the transmissibility of a virus, representing average number of secondary infectees generated by a primary infector[17]. For $R_0 > 1$, the number infected is likely to multiply, and for $R_0 < 1$, transmission is likely to die out[18]. We proxy daily SA by the satellite concentration of $NO_2$ (Sentinel-5P level-2 pollutant product)[19] because it is significanlty negatively associated with SA and available globally. The weather data are from the meteorolgcial reanalysis[20]. We quantify MOC through regionalized time fixed effects. Depening on the infectee records availability, we estimate 6,386 daily $R_0$ values from 3,315,748 confirmed cases of 213 countries or provinces/states from late January to mid April. Besides the overall contribution, we also compare the regional difference of the contribution, and the MOC evolvement. We finally estimate that compared to a hypothetical scenario with no restriction of SA, the global $R_0$ reduction contributed by the limiting SA from late January to mid April, and the consequent spared number of infectees.

## Results

Quantifying the overall contribution of SA and climateUsing all global records, we find that SA, Ultraviolet (UV), temperature (***T***), windspeed ($v$), and relative humidity (RH) significant affect $R_0$ ($R^2$=0.51 and all p<0.001, Table 2, Extended Data). Esimated by the product of the dependent-variable coefficient and value range over the response-variable value range, Table 1(a) shows strongly positive contribution of SA, moderately positive and negative contribution of temperature and UV, and weakly negative contribution of windspeed and RH. Therefore, SA restriction, UV illumination, high wind and high humidity could reduce the transmission. SA has a leading contribution of 18.3%±2.0% (95% Confidence Interval, CI). Sunlight kills the virus directly, increases cell cytotoxicity and modulates the biosynthesis of vitamin D which are essential to promote immune responses[21]. Higher relative humidity prevents viral particles traveling as far as it could in drier air[22].

We further control the regions to only include data in China and developed countries to confirm the contribution of SA and weather. The $R^2$ is increased to 0.61 and the SA contribution is increased to 23.2%±3.0% (95% CI), 4.9% higher than the global scale, whereas UV, wind, and RH become insignificant (Table *3*, Extended Data). The result might be explained by the fact that in more developed areas, people tend to stay indoors with a controlled climate most of the time[21]. This suggests that simply depending on favorable weather conditions, such as warm weather and high UV, has strong limitations in stopping the virus transmission in these areas.

### The nonmonotonic temperature contribution

We suspect a nonmonotonic contribution of temperature by restricting temperature to low or high (T<25 ℃ or T≥25 ℃) in the regresion because the coefficient of temperature changes from weakly positive to strongly negative (Table *4* vs. Table *5*, Extended Data). We hypothesize that the virus might have a viable temperature range. This finding agrees with a few previous studies on coronavirus[23] and

COVID-19[24,25], but disagrees with most studies claiming temperature has a decreasing[26,27], increasing[22] or no effect[28] on the transmissibility. However, we note that the sample size, with only 747 samples, is relatively small when restricting tempeture to high (T≥25°). More data in the incoming season are needed to confirm the temperature effects because most regions in the globe have not experienced a COVID-19 pandemic with high temperature.

## Global transmissibility reduction through restricting SA

We estimate the global decrease of $R_0$ due to the SA restriction as half monthly mean in Figure 1. Because China has the earliest outbreak since late January, earlier than the rest of the world, we provide the $R_0$ reduction from late January to Feburary only in China in Figure 1(a)-(c) and the $R_0$ reduction of the world from March to mid April in Figure 1(d)-(f). China imposed the strongest policy on restricting SA at an early stage, from Feburary to the first half of March, and saw the largest reduction of $R_0$. The major outbreak provinces of China had a $R_0$ reduction from 0.08 to 0.18 in the first half of Feburary, and from 0.02 to 0.12 in the latter half of Feburary and first half of March. China gradually lifted the restriction of SA since the latter half of March (Figure 1 *e* and *f*). The SA restriction in Europe came in the latter half of March, and the $R_0$ reduction was from 0.02 to 0.06 in the latter half of March and from 0.04 to 0.06 in the first half of April. The western US started SA strictions in the latter half of March, earlier than the rest of the US. The western US saw a reduction of $R_0$ from 0.02 to 0.08 from later March to April, and the eastern saw $R_0$ reduction from 0.04 to 0.10 in April.

Consequently, compared to a hypothetical scenario with no SA restriction during the community spread, we estimate in Figure 1(*g*) that China spared 40,964 (95% CI 31,463-51,470) infectees with 37,727 (95% CI, 28,925-47,488) in the Hubei Province, its outbreak center. Europe spared 174,494 (95% CI 139,202-210,841) infectees, where Germany, France, Spain, Italy, and United Kingdom spared 22,674 (95% CI 18,076-27,461), 24,463 (95%CI 19,527-29,541), 28,857 (95% CI 23,058-34,812),32,541 (95% CI 25,855-39,479), 18,580 (95% CI 14,864-22,386) respectively. The United States (US) spared 180,336 (95% CI 142,860-219,445) with 79,813 (95% CI 62,887-97,653) in New York State (The full table of spared infectees over time is available in the SI).

The number of spared infectees is a compound result of $R_0$ reduction and timing. Athough the $R_0$ reduction in Europe and US is significantly smaller than that in China, it spared significantly more infectees because of their higher number of infectors when the restriction of SA started. **Error! Reference source not found.** shows the regional spared infectees through SA restrictions in half-month intevals. Only China Mainland, Taiwan, South Korea, and Australia show a diminshed trend, as a result of early restrictions of SA, whereas the Europe and the US continue to see large number of spared infectees growing exponentially by the end of this analysis. The rest of the world, though having fewer number of spared infectees than Europe and US, show rapid growth. Maintaining restricted SA might still be a safer option for them. We suspect the lifting of SA restrictions to reopen the economy in these countries might pose great risk.

## Medical facility capacity and other control factors (MOC) evolvement

In Figure 3, the MOC is descreased the fastest in Australia and China to a value lower than or close to -1 eventually, indicating a resilient medical system or more effective control policy. The effectivness of the MOC in Australia might be explained by the adoption of additional control measures including mandatory quarantine of overseas arrivals and travel bans outside of Australia, and the relative small number of infectors in the beginning of the outbreak[29]. Although China was hit much harder in the begining, it survived the stress by imposing more strict control policies including the lockdown of the Hubei province, large-scale demobilization and mandatory quarantine of returning migrants (domestic) workers[29], and by promptly increasing medical facilities including temporary hospitals[30] and reinforced medical crews from other provinces to the outbreak center[31]. The MOC of Europe, US, and arctic countries starts from a high value and decreases rapidly, but has not reached a low value at the end of our analyses (above or close to zero), indicating that their medical facility capacity and control polices (other than restricting SA) is evolving rapidly while still under stress of the large number of infectees accumlated during an earlier phase of the outbreak. The MOC of Africa and Latin America, though started with a low number, does not decrease as rapidly as the rest of the world.

### Limitations

Our study has the following limitations. First, $R_0$, the reproductive number, is estimated from confirmed infectee records which can be inaccurate due to the lack of tested cases especially in the beginning of the outbreak and in regions with less testing capacity. Without globally distributed epidemic parameters of COVID-19 other than the confirmed cases, the adopted statistical estimation seems to be the only practical global solution (Method). Second, although the weather impacts are significant and are within expectation, the contribution of temperature, is subjected to revisitation after more countries experience the warmer weather. Finally, because the MOC is not parameterized but merely quantified, we cannot project MOC into the future. Therefore, our result should only be used to assess the contribution of each predictor, especially the restriction of socioeconomic activities, instead of predicting the total number of infectees in the future climate, or the end of the COVID-19 pandemic.

## Figure and Tables

Table 1. The contribution of predictors: (a) using all global (6,386) records, (b) using data only from China, and developed countries (4,371 records), and (c) using only records with and T<25 (5,565 records)

(a)

| Predictors | Contribution |
|---|---|
| SA | 18.3±3.5% |
| temperature | 8.0+2.6% |
| UV | -7.9±3.3% |
| windspeed | -5.7±1.8% |
| RH | 3.6%±1.9% |

(b)

| Predictors | Contribution |
|---|---|
| SA | 23.2±3.0% |
| temperature | 5.6±2.3% |

(c)

| Predictors | Contribution |
|---|---|
| SA | 21.4±3.0% |
| temperature | 5.4±2.0% |

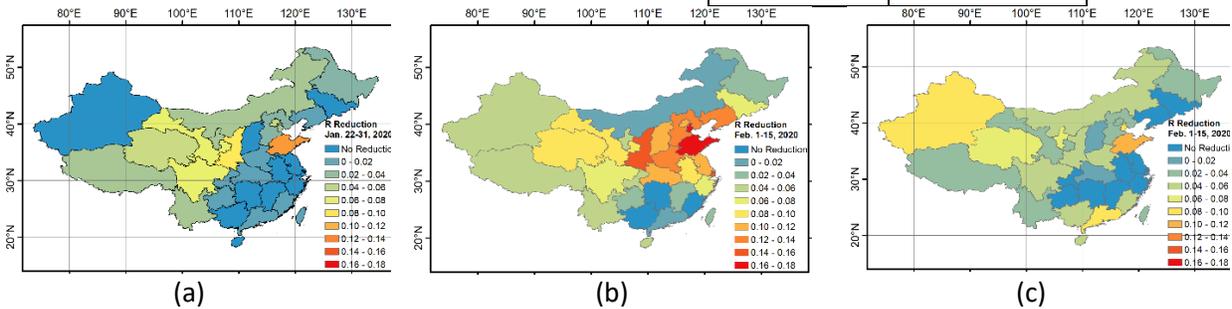

(a)        (b)        (c)

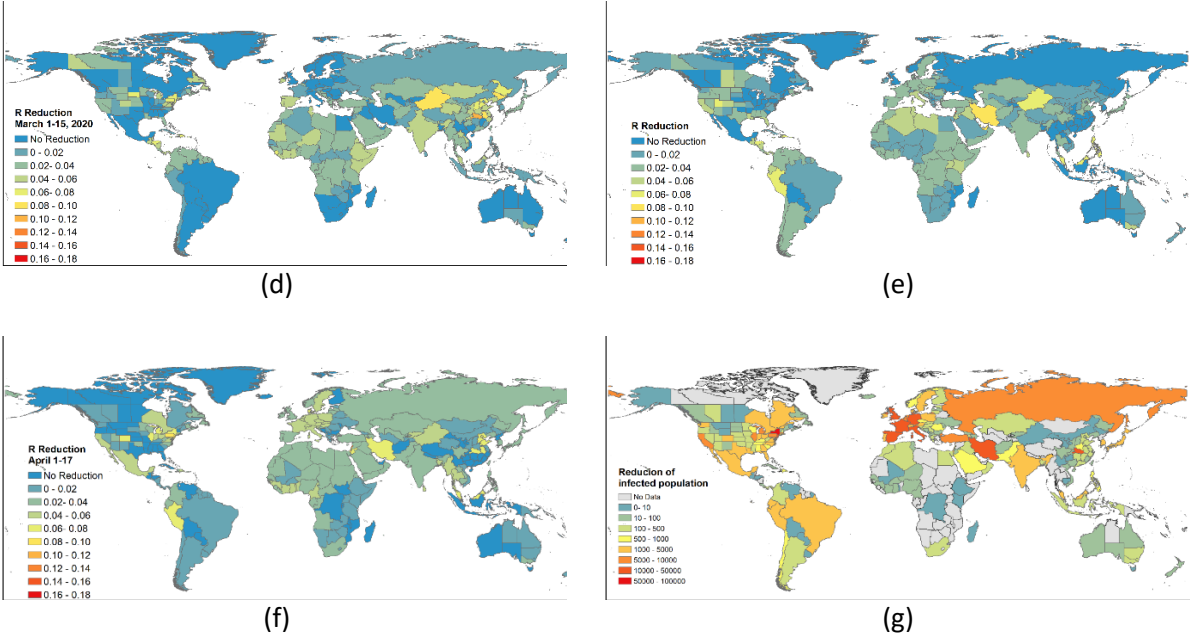

Figure 1. Global reduction of the transmission per administrative unit contributed by RSA, in terms of (a)-(f) half monthly mean reduction of the reproduction rate (R) from January 22$^{nd}$ to April 17, 2020, and (g) the total reduction of infected population. Since the community spread is only reported in China from before March, (a)-(c) do not include the rest of the world. In (g), the reduced population in administrative units without sufficient COVID-19 records for community spread cannot be estimated (marked as no data).

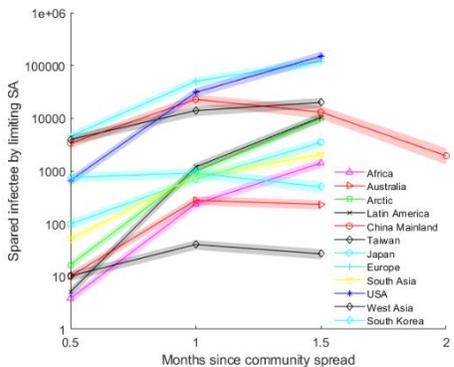

Figure 2. Regional reduced populations from infection through limiting SA (95% CI shaded area).

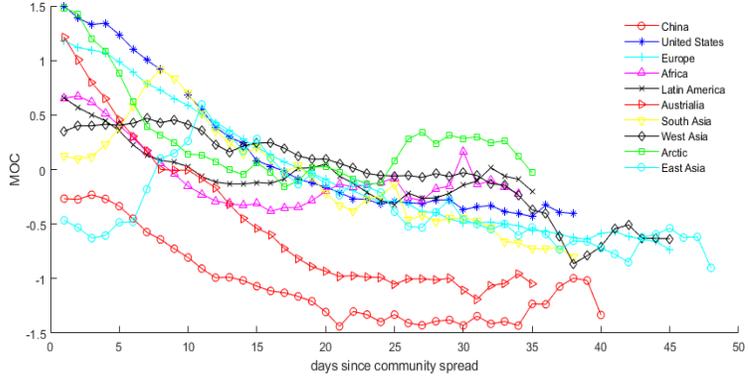

Figure 3. MOC contribution to the $R_0$.

# Methods

## The reproductive number estimation

Since except for the number of infectees and deaths, epidemic parameters of COVID-19 are not globally available, we employ the stochastic approach by [8,16] to estimate the reproductive number, $R_0$. $R_0$ of a given country or state/province at day $t$, $R_{0t}$, can be estimated from daily confirmed number of infectees (C), via an intermediate variable, daily number of new infectors (I). The conversion from $C_t$ to $I_t$ is given by equation (1),

$$I_t = \sum_{\tau=t+1}^{t+N} C_\tau lgn_{\tau-t} \quad (1)$$

where $lgn_t$ is probability of an infectee to get clinical confirmation $t$ days since being infected[16]. The conversion from $I_t$ to $R_{0t}$ is given by equation (2),

$$R_{0t} = I_t / \sum_{\tau=t-N}^{t-1} I_\tau g_{t-\tau} \quad (2)$$

where $g_t$ is the probability of an infector to infect other people after $t$ days since being infected[8].

## The contribution model and software

We fit a panel regression model with region-by-time fixed effects[3] (MOC) to quantify the contribution of SA and weather conditions to $R_0$, using the REGHDFE module in Stata software[4,32],

$$R_{0i,t} = \alpha + \beta_1 \ln(NO_2)_{i,t} + \beta_2 T_{i,t} + \beta_3 \ln(UV)_{i,t} + \beta_4 RH_{i,t} + \beta_5 v_{i,t} + \delta_{r,t} \quad (3)$$

where $NO_2$, $T$, $UV$, $RH$, $v$ stand for the total column of Nitrogen Dioxide concentration($mol/m^2$), air temperature (°C), ultraviolet intensity ($W/m^2$), relative humidity (%), and windspeed ($m/s$), and $\delta_{r,t}$ is the region-by-time fixed effects, with subscripts $i$ and $r$ standing for the response unit and region. A response unit is either a country or a province/state depending on the availability of COVID-19 records. For example, records are available in China, US, Canada, and Australia at province/state level while only available at country level in most other regions. $t$ stands for the days after community spread ($C_t>30$). The world is divided into 10 regions depending on the geographic proximity and data availability. They are, China, United States, Europe, Africa, Latin America, Australia (Australia and New Zealand), West Asia, South Asia, East Asia and Arctic (Canada and Russia).

The reduction of $R_{0i,t}$ by restricting SA can be estimated by the product of $\beta_1$ and the difference of between the $\ln(NO_2)_{i,t}$ under restriction and the $\ln(NO_2)_{ref}$ with no restriction which is averaged from the same period of the last year, i.e., 2019. Note that we only averaged the same period using the year of 2019 because Sentinel-5P data is only available after June 2018. Substituting the hypothetical $R_{0i,t}$ back to equation (2), we can estimate the number of infectees with no restricted SA.

## Data Source and Processing Method

$R_0$ is estimated from daily clinical confirmation of COVID-19 infectees assembled at Johns Hopkins University[1] in excel tables. The $NO_2$ concentration is retrieved from Sentinel-5P level 2 product[19] at daily intervals of 3.5 km ×7.5 km nadir resolution in netcdf4 format. Climate predictors are extracted from the ERA 5 atmospheric reanalyses[20] at hourly intervals and 0.25° ×0.25° grids in GRIB format. All climate predictors are averaged to daily mean and then averaged to the country or state/province weighted by population density from Gridded Population of the World (GPW)[33] in GeoTiff format. The $NO_2$ concentration is first spatially averaged to sub-level administrative areas of the response unit then weighted averaged by population density to the response unit.

## Extended Data

In this section, we provide the regression results mentioned in the main text to show the significance and value range each predictor.

Table 2. Regression result using the global 6,379 records. $R^2$=0.51, RMSE=0.63.

| Predictor | Coef. w. 95% CI | | | t | p>|t| | Variable | Min | Max |
|---|---|---|---|---|---|---|---|---|
| ln($NO_2$) | .2647 | .2143 | .3149 | 10.31 | 0.000 | R | 0 | 6.290 |
| T | .0087 | .0059 | .0115 | 6.13 | 0.000 | ln($NO_2$) | -10.60 | -6.252 |
| ln(UV) | -.1647 | -.2329 | -.0947 | -4.73 | 0.000 | T | -21.94 | 36.10 |
| $v$ | -.0327 | -.0430 | -.0223 | -6.17 | 0.000 | ln(UV) | 1.207 | 4.231 |
| RH | -.0025 | -.0038 | -.0011 | -3.55 | 0.000 | $v$ | .0090 | 11.03 |
| intercept | 4.6421 | 4.1925 | 5.0917 | 20.24 | 0.000 | RH | 8.153 | 98.47 |

Table 3. Regression result using the 4,371 records in developed countries and China. $R^2$=0.61, RMSE=0.58

| Predictor | Coef. w. 95% CI | | | t | p>|t| | Variable | Min | Max |
|---|---|---|---|---|---|---|---|---|
| ln($NO_2$) | .3536 | .3073 | .3998 | 11.13 | 0.000 | R | 0 | 6.290 |
| T | .0071 | .0042 | .0100 | 4.84 | 0.000 | ln($NO_2$) | -10.38 | -6.252 |
| ln(UV) | - | - | - | -1.18 | 0.236 | T | -21.94 | 27.75 |
| $v$ | - | - | - | 1.83 | 0.068 | | | |
| RH | - | - | - | -2.12 | 0.034 | | | |
| intercept | 4.6624 | 4.2237 | 5.1010 | 20.84 | 0.000 | | | |

Table 4. Regression result using 4,515 records with T<25 ℃. $R^2$=0.56, RMSE=0.60.

| Predictor | Coef. w. 95% CI | | | t | p>|t| | Variable | Min | Max |
|---|---|---|---|---|---|---|---|---|
| ln($NO_2$) | .3097 | .2658 | .3536 | 13.83 | 0.000 | R | 0 | 6.290 |
| T | .0072 | .0044 | .0099 | 5.06 | 0.000 | ln($NO_2$) | -10.60 | -6.252 |
| ln(UV) | - | - | - | -3.10 | 0.002 | T | -21.94 | 24.98 |
| $v$ | - | - | - | -1.11 | 0.265 | | | |
| RH | - | - | - | -2.35 | 0.019 | | | |
| intercept | 4.3164 | 3.8778 | 4.7569 | 162.16 | 0.000 | | | |

Table 5. Regression result using 774 records with T≥25 ℃. $R^2$=0.34, RMSE=0.76.

| Predictor | Coef. w. 95% CI | | | t | p>|t| |
|---|---|---|---|---|---|
| ln($NO_2$) | - | - | - | 2.50 | 0.013 |
| T | -.1712 | -.2146 | -.1279 | -7.75 | 0.000 |
| ln(UV) | - | - | - | -0.26 | 0.796 |
| $v$ | -.1397 | -.1709 | -.1086 | -8.80 | 0.000 |
| RH | -.0142 | -.0193 | -.0091 | -5.48 | 0.000 |
| intercept | 7.3858 | 5.9275 | 8.8441 | 9.94 | 0.000 |

# References

1	Dong, E., Du, H. & Gardner, L. An interactive web-based dashboard to track COVID-19 in real time. *The Lancet infectious diseases* (2020).
2	Report of the WHO-china joint mission on coronavirus disease 2019 (covid-19). <https://www.who.int/docs/default-source/coronaviruse/who-china-joint-mission-on-covid-19-final-report.pdf> (World Health Organization, Geneva, 2020).
3	Greene, W. H. in *Econometric analysis* (ed Donna Battista) Ch. 11 Models for panel data, 383-471 (Pearson Education Limited, 2003).
4	Correia, S. A feasible estimator for linear models with multi-way fixed effects. **working paper** (2016). <preprint at http://scorreia.com/research/hdfe.pdf>.
5	United Nations, *Country classification*, <https://www.un.org/development/desa/dpad/wp-content/uploads/sites/45/WESP2018_Annex.pdf> (2018).
6	Communications, D. o. (ed World Health Organization) 18 (2020).
7	Lipsitch, M. *Will COVID-19 go away on its own in warmer weather?*, <https://ccdd.hsph.harvard.edu/will-covid-19-go-away-on-its-own-in-warmer-weather/> (2020).
8	Flaxman, S. *et al.* Estimating the number of infections and the impact of non-pharmaceutical interventions on COVID-19 in European countries: technical description update. (2020).
9	Anderson, R. M., Heesterbeek, H., Klinkenberg, D. & Hollingsworth, T. D. J. T. L. How will country-based mitigation measures influence the course of the COVID-19 epidemic? *The Lancet* **395**, 931-934, doi:DOI: https://doi.org/10.1016/S0140-6736(20)30567-5 (2020).
10	Bedford, J. *et al.* COVID-19: towards controlling of a pandemic. *The Lancet* **395**, 1015-1018, doi:https://doi.org/10.1016/S0140-6736(20)30673-5 (2020).
11	Dehning, J. *et al.* Inferring change points in the spread of COVID-19 reveals the effectiveness of interventions. *Science*, doi:DOI: 10.1126/science.abb9789 (2020).
12	Lai, S. *et al.* Effect of non-pharmaceutical interventions to contain COVID-19 in China. *Nature* **in press**, doi:10.1038/s41586-020-2293-x (2020).
13	Li, Z., Shi, J. & Guo, H. in *Geoscience and Remote Sensing Symposium, 2002. IGARSS'02. 2002 IEEE International.* 3071-3073 (IEEE).
14	Yang, Z. *et al.* Modified SEIR and AI prediction of the epidemics trend of COVID-19 in China under public health interventions. **12**, 165 (2020).
15	Prem, K. *et al.* The effect of control strategies to reduce social mixing on outcomes of the COVID-19 epidemic in Wuhan, China: a modelling study. *The Lancet Public Health*, doi:https://doi.org/10.1016/S2468-2667(20)30073-6 (2020).
16	Lauer, S. A. *et al.* The incubation period of coronavirus disease 2019 (COVID-19) from publicly reported confirmed cases: estimation and application. *Annals of internal medicine* (2020).
17	Van den Driessche, P. & Watmough, J. in *Mathematical epidemiology* 159-178 (Springer, 2008).
18	Liu, Y., Gayle, A. A., Wilder-Smith, A. & Rocklöv, J. J. J. o. t. m. The reproductive number of COVID-19 is higher compared to SARS coronavirus. (2020).
19	Eskes, H. *et al.* Sentinel-5 precursor/TROPOMI Level 2 Product User Manual Nitrogendioxide. (Technical Report S5P-KNMI-L2-0021-MA, 2019).
20	Copernicus Climate Change Service (2017). *ERA5: Fifth generation of ECMWF atmospheric reanalyses of the global climate*, https://cds.climate.copernicus.eu/cdsapp#!/home.
21	Moriyama, M., Hugentobler, W. J. & Iwasaki, A. Seasonality of respiratory viral infections. *J Annual Review of Virology* **7** (2020).
22	Ma, Y. *et al.* Effects of temperature variation and humidity on the death of COVID-19 in Wuhan, China. *Science of The Total Environment*, 138226 (2020).

23  Chan, K. *et al.* The effects of temperature and relative humidity on the viability of the SARS coronavirus.  **2011** (2011).
24  Wang, M. *et al.* Temperature significant change COVID-19 Transmission in 429 cities.  (2020).
25  Ficetola, G. F. & Rubolini, D. J. m. Climate affects global patterns of COVID-19 early outbreak dynamics.  (2020).
26  Wang, J., Tang, K., Feng, K. & Lv, W. High temperature and high humidity reduce the transmission of COVID-19. *SSRN*, doi:Available at SSRN: http://dx.doi.org/10.2139/ssrn.3551767 (2020).
27  Qi, H. *et al.* COVID-19 transmission in Mainland China is associated with temperature and humidity: A time-series analysis. *Science of the Total Environment* **728**, 138778, doi:https://doi.org/10.1016/j.scitotenv.2020.138778 (2020).
28  Yao, Y. *et al.* No Association of COVID-19 transmission with temperature or UV radiation in Chinese cities. *European Respiratory Journal* **55** (2020).
29  International Monetary Fund (IMF), *Policy response to COVID-19*, <https://www.imf.org/en/Topics/imf-and-covid19/Policy-Responses-to-COVID-19> (2020).
30  Li, H., Zheng, S., Liu, F., Liu, W. & Zhao, R. Fighting against COVID-19: innovative strategies for clinical pharmacists. *Research in Social Administrative Pharmacy* **in press**, doi:https://doi.org/10.1016/j.sapharm.2020.04.003 (2020).
31  Chan, M. in *South China Morning Post*.
32  Correia, S. REGHDFE: Stata module to perform linear or instrumental-variable regression absorbing any number of high-dimensional fixed effects. *Statistical Software Components* **S457874**, doi:College Department of Economics (2014).
33  Center for International Earth Science Information Network - CIESIN - Columbia University (2018). *Gridded Population of the World, Version 4 (GPWv4): Population Density*.